# Sulfur and few-layer graphene interaction under thermal treatments

C. Bautista-Flores (1), J. S. Arellano-Peraza (2), R. Y. Sato-Berrú (3), E. Camps (4) and D. Mendoza (1)

((1) Instituto de Investigaciones en Materiales-UNAM, (2) Universidad Autónoma de México-Azcapotzalco, (3) Centro de Ciencias Aplicadas y Desarrollo Tecnológico-UNAM, (4) Instituto Nacional de Investigaciones Nucleares)

*claudiabautistaf@gmail.com **doroteo@unam.mx

In this work we study sulfur confined between two multilayer graphene films under thermal treatments by means of electrical and Raman spectroscopy characterization. We found some similarities between the electrical behavior and differential scanning calorimetry of sulfur immersed into nanoporous anodic alumina template during sulfur phase transitions, this suggests that sulfur has a different thermal behavior when it is in confining conditions compared to the free state. In addition, multilayer graphene was exposed to sulfur vapors at 500 °C to induce chemical reaction. This method effectively produced covalent bonding between sulfur and multilayer graphene and a p-type doping of about $2\times10^{13}$ cm$^{-2}$ in charge concentration. Finally, based on first principle calculations we speculate on the existence of a new bidimensional structure of sulfur.

## Introduction

Graphene is a very promising material for electronic applications because of its low sheet resistance, high optical transparency and mechanical properties. But the absence of a gap energy and a zero carrier density at Dirac points are in conflict with real applications [1] Therefore, a lot of effort has been focused to dope this material in order to improve its electronic properties. Heteroatom doping using nitrogen and boron has served to obtain n-type and p-type graphene respectively [2]. In this work we are mainly interested in graphene-sulfur system because of its amazing properties and potential applications in supercapacitors, Li-S batteries, catalyst supports, solid acids and many other areas [3].

Previously it has been found that sulfur modifies electrical properties of carbon-based materials, such as carbon nanotubes [4, 5] and graphite [6]. For graphene, it has been reported experimental and theoretically [7-9] that this material can act as a gas sensor [10], but S-doped graphene is much more reactive and can serve as a better sensor for polluting gases such as NO and $NO_2$ [7, 11] than pristine graphene. Also it was predicted that sulfur doping could open an energy band gap in graphene [8, 9]. On the other hand, Zhu and coworkers [12] found that sulfur dopants quench magnetic ordering in graphene and in general sulfur changed the magnetic behavior of their samples. Other experimental studies in graphite-sulfur composites have found superconducting behavior below 35K [13]. But this superconducting behavior is lost in monolayer graphene because layer-layer interactions in graphite play a key role [7, 14]. Also, confinement of sulfur in a nanoporous anodic alumina template [15, 16] have shown that the behavior of sulfur under this condition was different to that in bulk. These two last results suggest that sulfur confined between graphene layers can exhibit interesting properties. That is why the main purpose of our work is to confine sulfur (S) between two films of multilayer graphene (MG) and also we propose a simple and effective method to induce sulfur doping in graphene using thermal treatments.

## Experimental

Multilayer graphene was synthesized by chemical vapor deposition technique (CVD) using 3cmx1cm copper foils. The Cu foils were put into a quartz tube of a horizontal furnace which was heated from room temperature to 1000 °C with a hydrogen flow of 146 sccm. The system was maintained at these conditions during 90 minutes for annealing copper foils. Then, 40 sccm of methane passed through the quartz tube during 30 minutes at the same temperature. Finally methane flow was cut off and the furnace was cooled down to room temperature. In order to create a cross junction we cut 2mmx10mm ribbons of MG on copper foils. Then ferric nitrate solution was used to etch copper during 12 hours. MG ribbons were washed in de-ionized water to eliminate ferric nitrate solution residuals. One ribbon was translated to a glass substrate, when the sample was dried we deposited a sulfur film in the middle of the ribbon by spin coating using a sulfur and $CS_2$ solution (1 g of sulfur powder with 5 ml of $CS_2$, J. T. Baker 99.999% and Tecsiquim 99.9 % respectively). In order to cover with sulfur just an area of 2mmx2mm on the MG ribbon we used a wet paper mask. Finally another MG ribbon was put on top on S/MG making a cross junction as is shown in the inset of Figure 1 a), we called MG/S/MG to the final sample. Our MG films are a mixture of crystals with a different number of graphene layers, from one to about 5 layers.

For electrical measurements MG/S/MG samples were mounted on a sample holder which is introduced into a sealed chamber. Silver paste was used to make electrical connections, in a pair of MG ends an electrical current was passed and the voltage drop was measured in the opposite arms in the cross junction configuration, such as is illustrated in the inset of Figure 1 a). In an argon atmosphere, voltage drop as a function of temperature was monitored during two cycles, from room temperature to 200 °C and back to room temperature.

For Raman characterization we used a Nicolet Almega XR spectrometer and 532 nm of laser excitation, for these measurements samples were translated on silicon with 306 nm of silicon dioxide substrates. It should be noted that for all Raman characterization presented in this work, only graphene zones were selected on the MG film. A Linkam-THMS600 cell for Raman characterization at different temperatures in an Argon atmosphere was used. We found that when these measurements were made above 100 °C, most of the sulfur evaporates; consequently we used other method to induce doping in graphene with sulfur, which consists on using sulfur in vapor phase. Samples were put into a quartz tube inside a tubular furnace. The system was heated from room temperature to 500 °C in an argon atmosphere, sulfur powder was put into the same tube but in a position where the temperature was around 200 °C. At this temperature sulfur can evaporate and vapor flows along the tube and finally react with MG at 500 °C. We called MG+S to the resulting sample. Then Raman spectroscopy was used again to analyze these samples and compared with the cross junction samples (MG/S/MG). To determine the incorporation of sulfur into MG structure we used a K-Alpha Thermo Fisher Scientific equipment for XPS characterization.

## Results and discussion

Electrical resistance of the cross junction as a function of temperature is shown in Figure 1 a). Here we compare samples with (MG/S/MG) and without (MG/MG) sulfur between MG films. In samples with sulfur there are some notable changes at around the following temperatures: 92 °C, 112 °C, 123 °C and 167 °C. These changes are related with phase transitions of sulfur or interactions between sulfur and MG because they do not appear in samples without sulfur. It is known that during heating bulk sulfur suffers different transitions: from room temperature to 95 °C the thermodynamically stable phase is $S_\alpha$, which consists of an eight-atomic cyclic molecule (cyclo-octa-S). At 95 °C ($T_{\alpha-\beta}$) α-sulfur transforms into monoclinic β-sulfur [15, 17]; this phase is stable from 95 °C to about the melting point at 119 °C, labeled by $T_m$ in Figure 1 b). The other important change in sulfur is the so called λ transition at 159 °C ($T_\lambda$). After melting point viscosity decreases with temperature but there is a dramatic increase of it with a maximum at 159 °C, beyond this temperature viscosity decreases again [18].

For comparison, in Figure 1 b) differential scanning calorimetry of bulk sulfur (S) and sulfur immersed in a nanoporous anodic alumina template (NAA) are presented [15]. Also in this figure normalized resistance of Figure 1 a) is shown in order to compare with transition temperatures of sulfur. In this figure three important temperatures of sulfur are shown, $T_{\alpha-\beta}$, $T_m$ and $T_\lambda$. Carvajal and coauthors [15] found a different behavior of heat flow of sulfur when it is confined into the nanopores (labeled by NAA+S). In our case, sulfur is also confined between two layers of MG, if we compare heat flow of NAA+S with resistance of the cross junction of MG/S/MG we can observe that they present some peaks at the same temperatures. These observations suggest that the MG/S/MG arrangement is useful to monitor transitions in confined sulfur and possibly some characteristic temperatures of graphene-sulfur interactions.

On the other hand, Raman spectroscopy has proven to be one of the most complete characterization tool for graphene [19]. With this spectroscopy technique it is possible to obtain information of the number of graphene layers, disorder, doping, strain/stress, edges and more. We monitor sulfur-carbon interaction from room temperature to 100 °C in a MG/S/MG sample using in situ Raman spectroscopy, results are presented in Figure 1 c). As is shown in that figure, during heating Raman signal of sulfur was decreasing from 20 °C up to 75 °C, which may indicate that sulfur evaporates or is reacting with MG. At 100 °C sulfur signal is very small in comparison to the initial one (at 20 °C). The most important observation with Raman in situ at 100 °C is that the zone between D and G Raman peaks of graphene is modified, a very notable peak appeared and it is indicated with a green arrow in Figure 1 c). We tried to heat samples up to 200 °C but the rapid evaporation of sulfur did not permit us to monitor its Raman spectrum, as is shown in Figure 1 d) where Raman signal of sulfur at 200 °C and at room temperatures disappeared.

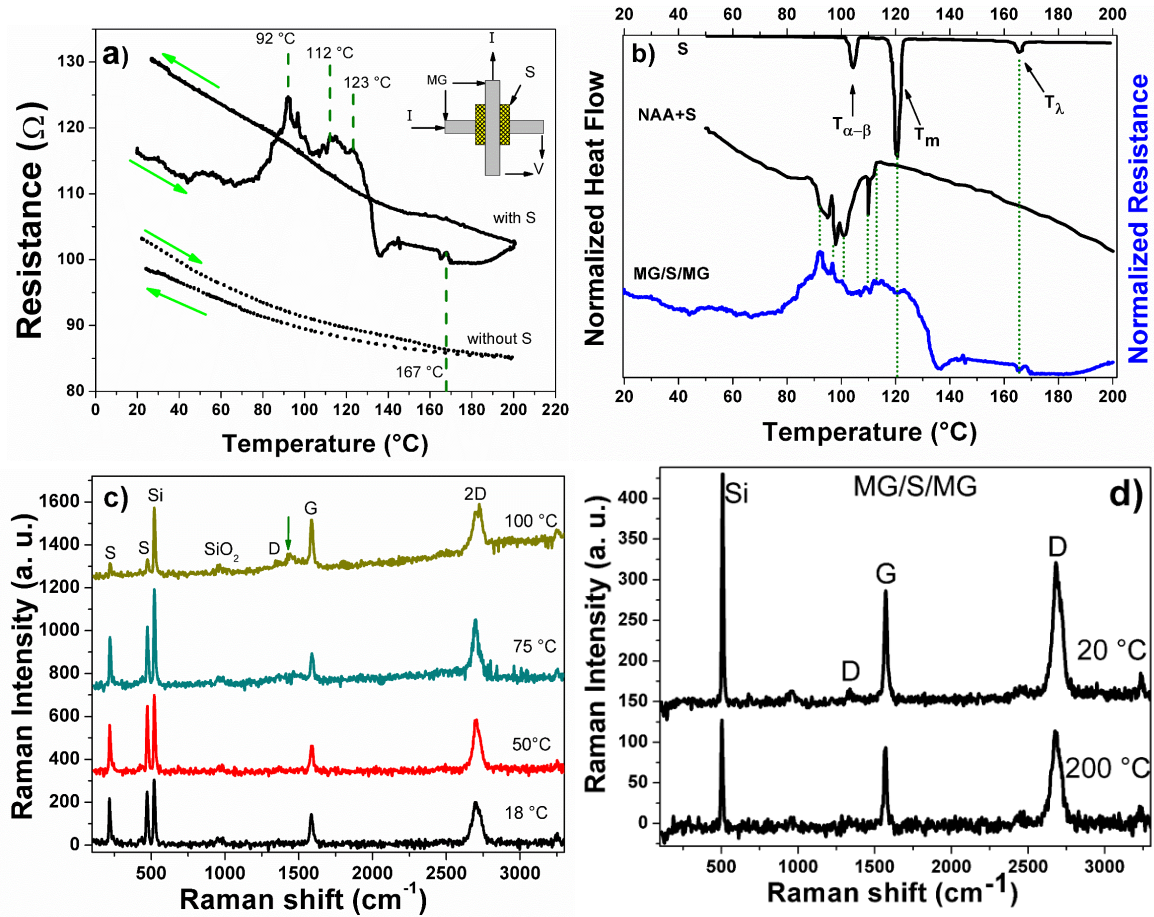

Figure 1: a) Resistance as a function of temperature of a cross junction with and without sulfur in the junction. b) Differential scanning calorimetry of bulk sulfur (S), sulfur immersed in a nanoporous anodic alumina (NAA+S) template (adapted from Ref. 15) and resistance of MG/S/MG as a function of temperature, the same sample in a); c) Raman in situ of a MG/S/MG on a silicon substrate, from room temperature to 100 °C, ; d) Raman spectra of MG/S/MG at 200 °C and back to 20 °C, in both sulfur Raman signal does not appear, which means that it has been evaporated.

We propose that in the sandwiched configuration (MG/S/MG) eventually a bi-dimensional configuration of sulfur atoms may be stable between MG; such as has been reported for new water structures between graphene layers [20, 21]. First principle calculations using quantum espresso computer code [22] show that when MG layers are separated by $d = 5.2818$ Å the $S_8$ molecular ring is broken and the atoms form a layer of $S_x$ chains in the middle plane between the graphene layers; these results are shown in Figure 2. In Figure 2 a) the initial state of the system is shown, in Figure b) and c) the final state is presented when the $S_8$ ring is broken, lateral and upper view respectively. In these calculations we have used $S_8$ rings because this is the unit molecule for the most stable sulfur structure at ambient conditions. To search on the origin of the Raman peak between D and G bands shown in Figure 1 c), and because of the rapid sulfur evaporation, we proceed to study the interaction of sulfur vapors and graphene at higher temperatures, namely 500 °C. A schematic representation of this method is shown in the inset of Figure 3 a). Results of these thermal treatments with sulfur are presented in Figure 3 a), b) and d). The zone between D and G peak in Figure 3 b) changes dramatically (similar to spectrum at 100 °C in Figure 1 c)) and it is very different to other sulfur-graphene reports [23-26]. This zone, expanded in Figure 3 b, is composed by peaks at 1328 cm$^{-1}$, 1436 cm$^{-1}$, 1530 cm$^{-1}$ and 1580 cm$^{-1}$. As we know, peaks at 1328 cm$^{-1}$ and 1580 cm$^{-1}$ are D and G peaks of graphene respectively [19, 27, 28]. We propose that the origin of the peak at 1530 cm$^{-1}$ is due to C-S vibration, such as the $\nu_3$ Raman mode that appears in liquid $CS_2$, as is shown in the inset of Figure 3c). In this figure we can notice that Raman spectrum of sulfur does not show a Raman signal in the region of $\nu_3$ Raman mode of $CS_2$. At this stage we do not know the origin of the peak centered around 1436 cm$^{-1}$. In Figure 3d) the position of both G (Pos(G)) and 2D (Pos(2D)) Raman bands of graphene for many samples are shown, which upshift up to 23 cm$^{-1}$ and 29 cm$^{-1}$ respectively. We adjudicated these changes to p-type doping in graphene produced by sulfur. Using the average of Pos(G) we obtain a charge concentration [29-32] of around $2 \times 10^{13}$ cm$^{-2}$; which is a considerable doping effect produced by sulfur.

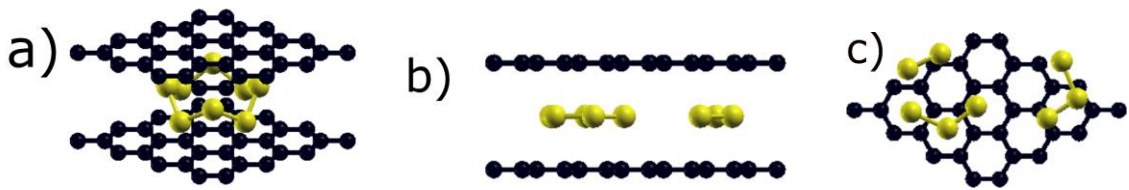

Figure 2: a) Initial state of $S_8$ ring between two layers of graphene; b) Final state of a), here the $S_8$ ring is broken; c) is an upper view of b), in this figure the top layer of graphene has been omitted for a better view.

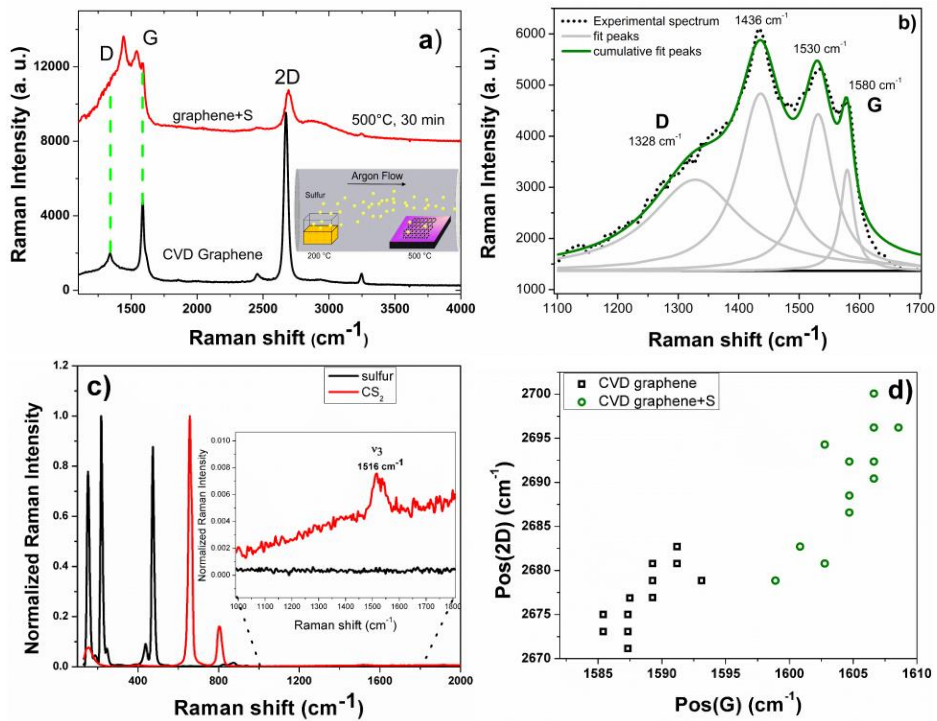

Figure 3. a) Raman spectra of CVD graphene and graphene after 500 °C thermal treatment with sulfur vapor (graphene+S), in the inset an schematic representation of thermal treatments with sulfur is shown; b) fitting peaks of the D-G zone of the Raman spectrum of graphene+S; c) Raman spectrum of sulfur powder and liquid $CS_2$, the inset in this figure shows the $\nu_3$ Raman mode of $CS_2$ which corresponds to C-S vibrations; d) Pos(2D) as a function of Pos(G) of graphene before and after thermal treatments with sulfur.

We believe that part of the observed changes in the D-G zone is due to defects and/or disorder in MG structure produced by the heating of the sample, such as has been reported in other work [33] but also by C-S bonding out of the graphene plane. Our calculations show that, when we have a monolayer graphene with two vacancies located in opposite vertices of and hexagon, a sulfur atom bonds out of the graphene plane, breaking the ideal bidimensionality (see Figure 4a). But it can also occur that two nearer vacancies are saturated by one sulfur atom located almost in the graphene plane, such as is observed in Figure 4b). Other possible configuration is presented in Figure S1 in Supplementary Information.

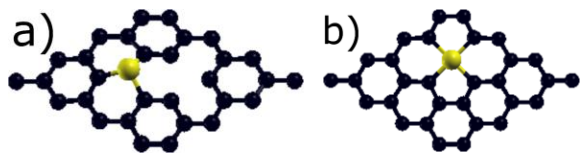

Figure 4: a) Two carbon vacancies located in opposite vertices and one sulfur atom adsorbed on one of this carbon vacancies, this bonding is located out of the graphene plane; b) one sulfur atom forming bonding with carbon atoms on a double vacancy practically in the graphene plane.

On the other side, XPS characterization was used to probe the existence of sulfur in the MG structure, see Figure 5 a). High resolution XPS of C1s in Figure 5 b) shows that the most intense component is at 284.3 eV (C=C), this means that $sp^2$ hybridization is the most important in our samples [24]. In Figure 5 c), high resolution S2p XPS exhibits two main peaks at 163.7 eV (C-$S_x$-C, x=1 or 2) and 165.1 eV (C=S) [34, 35]. The peak at 161.6 eV can be assigned to the reduced sulfur moieties -SH and peaks at 167.8 eV and 169.5 eV to the oxidized sulfur moieties (-$SO_x$) [34]. Therefore Raman and XPS analysis shows that sulfur incorporates successfully into MG lattice.

Finally, by first principle calculations using quantum espresso computer code [22], we have explored the possibility of the existence of a new form of sulfur; which we denominate "sulfurene", and it is presented schematically in the inset of Figure 6a. The energy of this system as a function of S-S distance is shown in Figure 6 a). This is a bidimensional hexagonal and planar structure (similar to graphene) with a S-S equilibrium distance of 2.2649 Å and a lattice parameter of 3.9229 Å. The electronic density of states of this structure, Figure 6 b), exhibits the Fermi energy at -3.2517 eV; which means that this sulfur structure has a metallic character. In Figure S2 we show a calculated Raman spectrum of this structure. Experimentally a linear structure of sulfur with dimensions within the range of 2-5 nm has been reported on S-doped graphene [24]. Sulfur atoms formed linear nanodomains which act as bridges to connect carbons and finally form the S-graphene network. In other report on the S-graphene system, an orthorhombic $CS_2$ (honeycomb lattice) structure has been found [11, 26]. But a bidimensional and hexagonal sulfur structure has not been reported yet. We believe that one possibility to have this new sulfur structure is that under special conditions it could be hosted in graphitic structures, such as in intercalation systems. We speculate that the superconducting traces observed in the graphite-sulfur system [13] might be related to this sulfur structure because the electronic properties of metallic sulfurene may change under confinement between graphene layers. On the other side, it has been reported that sulfur becomes superconducting under high pressures [36] and confinement between graphene layers may simulate this condition which may favor the possibility of superconductivity.

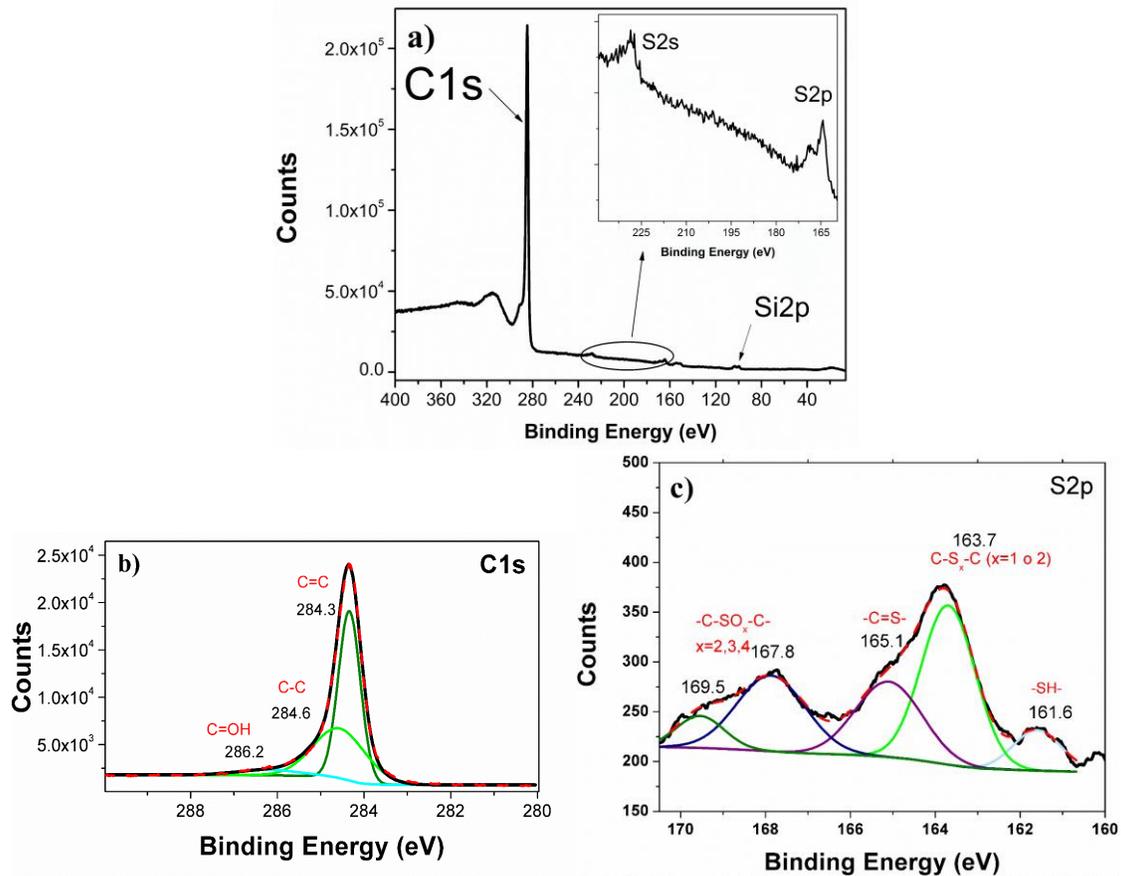

Figure 5: a) XPS full spectrum of graphene+S; High resolution XPS of b) C1s and c) S2p.

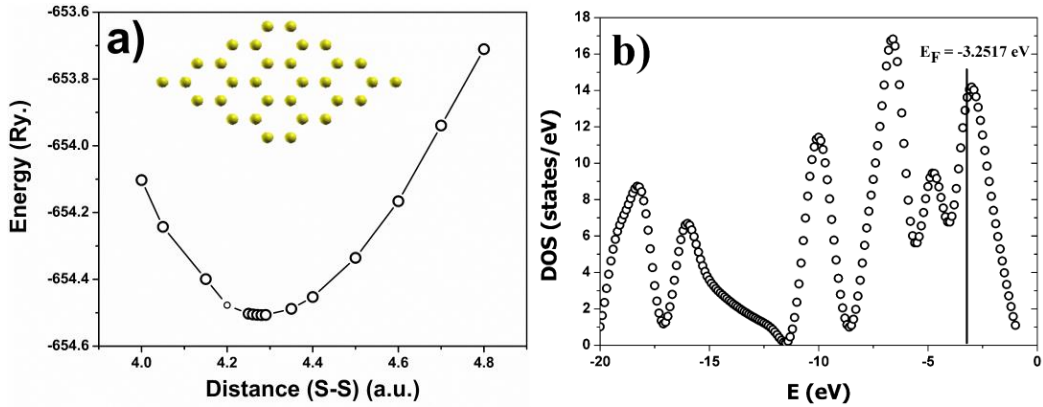

Figure 6: a) Energy of sulfurene as a function of S-S distance, this bidimensional structure is shown in the inset; b) Density of states of sulfurene, the Fermi energy of this material is located at -3.2517 eV.

## Conclusions

Using a cross junction of multilayer graphene films with sandwiched sulfur in the intersection it was possible to study sulfur under confinement situation. We found that using sulfur vapor to induce chemical reaction between sulfur and multilayer graphene, the Raman spectrum of our graphene-sulfur material is completely different in comparison with other works. Our method effectively incorporates sulfur into multilayer graphene lattice and charge concentration of about $2 \times 10^{13}$ cm$^{-2}$ was induced. Finally, using first principle calculations we propose the existence of a new bidimensional structure of sulfur similar to graphene.

## Acknowledgements

Authors want to thank to Laboratorio Universitario de Caracterización Espectroscópica (LUCE) at CCADET-UNAM for facilities in Raman characterization and Laboratorio de Supercómputo y Visualización en Paralelo at the Universidad Autónoma Metropolitana-Iztapalapa.